\documentstyle[12pt,aasms4]{article}

\lefthead{Bekki Kenji,  Yasuhiro Shioya,  \& Ichi Tanaka}
\righthead{Morphological and photometric evolution}

\begin{document}
\title{Morphological and photometric evolution of ultra-luminous infrared 
galaxies: Nature  of faint  SCUBA sources}

\author{Kenji Bekki, Yasuhiro Shioya\altaffilmark{1}, and Ichi Tanaka} 
\affil{Astronomical Institute, Tohoku University, Sendai, 980-8578, Japan} 

\altaffiltext{1}{Center for Interdisciplinary Research, Tohoku University, 
Sendai, 980-8578, Japan}

\begin{abstract}

 We investigate   when and how a dusty starburst galaxy merger
 can be heavily obscured by  dust and consequently becomes an  ultra-luminous
 infrared  galaxy (ULIRG), based on numerical simulations of chemodynamical and photometric
 evolution of dusty gas-rich major galaxy mergers.
 We found that a major galaxy  merger is more likely to become
 an ULIRG preferentially in  the merger late phase, when the two disks 
 become very close and the very high-density dusty gas can
 obscure heavily  the central secondary massive starburst.
 We furthermore show how the optical and near-infrared morphology of
 a simulated ULIRG at intermediate (z=0.4) and high redshift (z=$1 - 2$) 
 can be observed by the Hubble Space Telescope (HST)
 in order to present a plausible explanation for the origin
 of some host galaxies  of the faint SCUBA sources  recently observed by Smail et al. (1998).
 The results of our numerical simulations imply  that 
 some SCUBA sources  with apparently faint and compact HST morphology 
 can be higher redshift dust-enshrouded starburst mergers with their outer low surface
 brightness tidal features hardly detectable   in the present optics
 of the HST.

\end{abstract}


\section{Introduction}

 Recent observational studies with the Sub-millimeter Common-User Bolometer Array
 (SCUBA) (Holland et al. 1999) on the James Clerk Maxwell Telescope have revealed
 possible candidates of heavily dust-enshrouded starburst galaxies at intermediate
 and high redshift, which could be counterparts of low redshift  
 ultra-luminous infrared  galaxies (Smail, Ivison, \& Blain 1997;
 Hughes et al. 1998; Smail et al. 1998; Barger et al. 1999; Lilly et al.  1999).
 Although optical morphology of these sub-millimeter galactic sources should be treated with
 caution owing to the absence of high resolution sub-mm imaging capability (Richards 1999), 
 more than 50 \% of those are  suggested  to 
 show the indication of galactic interaction
 and merging (Smail et al. 1998).
 The huge amount of molecular gas  ($\sim 5 \times$ $10^{10}$ $ \rm M_{\odot}$) derived
 from CO emission and the inferred large star formation rate ($\sim {10}^{3} {\rm M}_{\odot} {\rm yr}^{-1}$)
 in a few sub-millimeter extragalactic sources (e.g. SMMJ14011+0252) are suggested to be
 consistent with the  scenario that the formation
 of some intermediate and high redshift elliptical
 galaxies is  associated with  galaxy mergers with their starburst activities heavily hidden
 by dust extinction (Frayer et al. 1998; Frayer et al. 1999).
 Furthermore observational studies of an extremely red object ERO J 164502-4626.4 (HR 10)
 with the redshift of 1.44 by the Hubble Space Telescope and the SCUBA have
 found that this high redshift galaxy  is also a dust-enshrouded
 starburst galaxy with clear indication of galaxy merging/interaction 
 (Graham \& Dey 1996; Cimatti et al. 1998; Dey et al. 1998).


 Although
 low redshift ultra-luminous infrared  galaxies (ULIRG) are generally considered to be ongoing
 galaxy mergers with the triggered  prominent nuclear activities (starburst or AGN)
 heavily obscured  by dust (Sanders et al. 1988; Sanders \& Mirabel 1996),
the origin of the faint SCUBA sources is not so clearly understood. 
In this Letter, we numerically investigate both morphological  and photometric
properties of dusty starburst galaxy mergers
in an explicitly self-consistent manner and thereby
demonstrate when and how a gas-rich merger  between two spirals becomes an
ULIRG.
We particularly discuss the origin of
very faint and compact HST optical morphology of  possible host galaxies of some faint SCUBA sources recently
revealed by Smail et al. (1998) by
demonstrating  how intermediate and high redshift dust-enshrouded starburst
galaxies can be seen by the HST WFPC2 and NICMOS (NIC2).

\section{Model}

The most remarkable difference in investigating photometric evolution
of dusty galaxies between the present model and previous one-zone models
(e.g., Mazzei, Xu, \& De Zotti 1992; Franceschini et al. 1994;
Guiderdoni et al. 1998) is that we derive spectral energy distribution (SED) of galaxies
based on the result of numerical simulations that can follow 
both  dynamical and chemical evolution of galaxies.
The numerical techniques for solving galactic chemodynamical and photometric evolution
are given in  Bekki \& Shioya (1999a,b),
and accordingly we describe only
briefly the present numerical model of dusty starburst galaxy mergers here. 
 We construct  models of galaxy mergers between gas-rich 
 disk galaxies with equal mass by using Fall-Efstathiou model (1980).
 The total mass and the size of a progenitor disk are 6.0 $\times$ $10^{10}$ $ \rm M_{\odot}$ 
 and 17.5 kpc,  
 respectively. 
  The collisional and dissipative nature 
  of the interstellar medium with the initial total mass
  equal to  $3.0 \times$ $10^{10}$ $ \rm M_{\odot}$  is  
 modeled by the sticky particle method
  (\cite{sch81}).
  Star formation 
  is modeled by converting  the collisional
  gas particles
  into  collisionless new stellar particles according to 
  the Schmidt law (Schmidt 1959)
  with the exponent of  2.0.
 Chemical enrichment through star formation during galaxy merging
is assumed to proceed both locally and instantaneously in the present study.
The fraction of gas returned to interstellar medium in each stellar particle
and the chemical yield
are 0.2 and 0.03, respectively.
Initial metallicity $Z_{\ast}$ for each stellar and gaseous
particle in a given galactic radius  $R$  (kpc) from the center
of a disk is given 
according to the observed relation $Z_{\ast} = 0.06 \times {10}^{-0.197 \times (R/3.5)}$
of typical late-type disk galaxies (e.g., Zaritsky, Kennicutt, \& Huchra 1994).
Mean ages of old stellar components
initially in  a merger progenitor   disk at the redshift z=0.4, 1.0, and 1.5 are
assumed to be 7.14, 3.80, and 2.46 Gyr, respectively. 
 The orbital plane of a  galaxy merger is assumed to be the same as $x$-$y$ plane
 and the initial distance between the center of mass of merger progenitor
 disks is 140 kpc.
 Two disks in the merger are assumed to encounter each other parabolically
 with the pericentric distance of 17.5 kpc.


It is assumed in the present study
that the SED of a model galaxy is 
a sum of the SED of  stellar particles. 
We first calculate dust extinction of star light for $each$  stellar particle
and dust temperature for $each$ gaseous particle, based on the three dimensional
spacial distribution of stellar and gaseous particles.
We  then  sum each stellar particle's  SED corrected by dust extinction and the
dust re-emission of each gaseous particle.
The method to derive the dust extinction and re-emission for each particle 
is described as follows.
Absorption of star light of a stellar particle is modeled according
to the following reddening formulation (Black  1987; Mazzei et al. 1992); 
$E(B-V)=N(\rm H)/4.77 \times {10}^{21} {\rm cm}^{-2} \times ({\it Z}_{\rm g}/0.02)$,
where $N(\rm H)$ and ${Z}_{\rm g}$ are gaseous column density and gaseous metallicity, respectively.
Using the derived redding $E(B-V)$ and extinction law by Cardelli et al. (1989),
we adopt the so-called screen model and  calculate the absorption of the stellar particle.
By  assuming  the modified black body radiation 
with  the emissivity ($\epsilon$) law $\epsilon \propto {\nu}^{2}$, 
we  determine the dust temperature of a gas particle such  that
total energy flux of dust absorption is equal to that of dust re-emission. 
We here do not include the albedo  for dust grains, which means that only stars and new stars
heat dust. 
For calculating the SED
of each stellar particle with  a given age and metallicity, 
we use the 
spectral library GISSEL96 which is the  latest version of  Bruzual \& Charlot (1993).
 Using  the derived SED, 
 we investigate how dust-enshrouded starburst galaxy mergers
 at z=0.4, 1.0, and 1.5 can be seen in the HST. 
 The method to construct the  synthesized HST images of galactic morphology
 in the present study is basically the same as that described by Mihos (1995).
 In the followings, the cosmological parameters
 $\rm H_0$ and $\rm q_0$ are  set to be 50 km s$^{-1}$ Mpc$^{-1}$ and 0.5
 respectively.

\placefigure{fig-1}
\placefigure{fig-2}
\placefigure{fig-3}

\section{Result}

Figure 1 and 2 describe the time evolution of star formation rate and the rest-frame SED 
in  a dusty gas-rich major galaxy merger at z=1.0 and the mass distribution
of stellar and gaseous components of the merger at the  epoch
of the maximum secondary starburst,  $T = 1.3$ Gyr, respectively.
From now on,  the time $T$ represents  the time that has elapsed since the two disks
begin to merge at each redshift.  
As is shown in 
Figure 1,  star formation rate in the present  prograde-retrograde
merger becomes  maximum ($ \sim  378  \rm M_{\odot}/ yr$) at $T = 1.3$ Gyr,
when two disks of the merger  become very close to suffer from final violent relaxation. 
These results are qualitatively consistent with those in Mihos \& Hernquist (1996). 
The SED at $T = 1.3$ Gyr in Figure 1 
is rather  similar to the observed SED of typical ULIRGs,
which implies that the present dusty starburst merger model can be 
observed as an ULIRG in the late of galaxy merging.
The SED of a dusty merger depends on encounter parameters in such a way  that 
a prograde-retrograde merger shows stronger infrared re-emission in the merger
late phase than prograde-prograde one (Bekki \& Shioya 1999b).

Figure 2 shows that both new stars and gas are more centrally concentrated 
than old stellar components,
which means that star light from new stars formed by secondary
massive starburst are more heavily obscured  by dusty interstellar medium
than that of old stellar components.
These  results in Figure 1 and 2 clearly demonstrate that starburst population in the merger
is very heavily obscured by dusty interstellar medium  ($A_{V} \sim 2.46$ mag) 
and consequently  the dust re-emission
in far-infrared  ranges 
becomes very strong at the maximum starburst
($L_{IR}=1.59 \times 10^{12} \rm L_{\odot}$). 
The reasons  for this heavy dust extinction are  firstly that 
column density of dusty interstellar medium  becomes
extremely high owing to the strong central accumulation of gas 
and secondly  that the central gaseous metallicity, which is a measure
of total amount of dust,
also becomes rather large ($\sim 0.04$)  because of efficient and rapid chemical evolution
in galaxy merging with secondary starburst.
The galaxy merger 0.05 Gyr before and after the maximum starburst  does not
show the SED characteristics of ULIRGs,  
which suggests  that the time-scale during which the  merger remnant 
can be observed as an ULIRG is very short (less than  0.1 Gyr).
Owing to the heavy dust extinction, the merger shows very red rest-frame color $V-I$=0.97 for
a starburst galaxy.

Figure 3  shows  how the optical and near-infrared morphology
of this ULIRG  merger at  $T = 1.3$ Gyr can be observed  by the HST WFPC2 and NICMOS
if it locates
at the redshift z=0.4, 1.0, and 1.5 (See  also Figure 2  for the raw image of the merger). 
Outer diffuse  large arc-like structure composed mainly of old stars, which is discernibly seen in the upper left
part of Figure 2,  can not be seen even in the merger at z=0.4.
Tidal feature indicative of  major galaxy merging can be  discernibly seen
at z=0.4 and can not be seen at all at z=1.0 and 1.5. 
The faint and compact morphology of the present merger remnant at  z=1.0 and 1.5 
is similar to some  possible faint optical counterparts of
the first SCUBA sources observed by Smail et al. (1998).
We furthermore stress  that  it is hard to observe the tidal feature even in  the  HST image with the exposure time
equal to $1.24  \times {10}^{5}$ sec corresponding to that for the Hubble Deep Field (HDF). 
These  results imply  that 
it is  not so easy task to clarify the origin of high redshift ULIRGs ($\rm z > 1$) solely by
optical and near-infrared HST morphological studies; If
the formation of high redshift ULIRGs is  closely associated with major galaxy merging,
even the deepest existing HST surveys
have some difficulties in proving that.
The 850 $ \rm \mu m$  sub-millimeter energy flux of the ULIRG merger is found to be
10.85 mJy for z=1.0 and 2.36 mJy for z=1.5.
Considering the flux value of the merger at  z=1.0 and 1.5 
exceeding the  observed SCUBA flux $2.1 -  7.0$ mJy
by Hughes et al. (1998) and $3.3 -  4.6$ by Barger et al. (1998) 
and faint compact morphology of  the merger,
a high redshift dust-enshrouded  starburst merger (z$\sim$1.0) can be observed as 
a very  compact SCUBA source.
Furthermore the apparently compact remnant shows very red $R-K$ color (4.70 mag for z=1.0 and 4.86
for z=1.5) for its $K$ band magnitude  (21.26 mag for z=1.0 and  and 22.42 for z=1.5),
which implies an evolutionary link between dusty mergers and the high redshift extremely red objects (EROs).
The present model  describes the detectability of only $one$ merger model with the initial disk mass 
 $M_{\rm d}$ = 6.0 $\times$ $10^{10}$ $ \rm M_{\odot}$ in the $present$ optics of the HST.
Accordingly we must  stress that the detectability of tidal feature  
of a  high redshift ULIRG  in the  HST can depend on mass and luminosity of a merger
and furthermore that  the  future improved HST Advanced Camera for Surveys (ACS)
is expected to reveal the more detailed morphology of a high redshift ULIRG.

\section{Discussion and conclusion}

 Our numerical results have demonstrated that a dust-enshrouded starburst merger
 can be observed as a SCUBA source with faint and compact morphology at high redshift.
 The present results furthermore provide  clues to the understanding
 of the evolutionary link between the extremely red objects `ERO' (e.g., Elston, Rieke, \& Rieke 1988)
 and the present day elliptical galaxies.
 Recent observational studies have revealed that ERO J 164502+4626.4 (HR 10)
 is not a  red elliptical galaxy passively evolving at high redshift 
 but a dust-enshrouded starburst galaxy at  redshift z=1.44 (Cimatti et al. 1998; Dey et al. 1998).
 Peculiar morphological properties in rest-frame far-red and near-UV range,
 the large 850 $ \rm \mu m$ energy flux, 
 and the inferred huge star formation rate 
 in this HR 10 are qualitatively consistent  with the present numerical
 results of dusty starburst mergers, which implies 
 that $some$  EROs are dusty starburst major mergers and thus  
 evolve eventually into the present-day elliptical galaxies.
 Then, what fraction of the present-day ellipticals have  experienced
 the past dusty ERO phase ? 
 Considering both the very short time-scale (less than 0.1 Gyr
 in the present study) during which dusty mergers can
 be identified as EROs and the high surface density
 of EROs (0.14 ${\rm arcmin}^{-2}$ for EROs with
 $R-K \ge 6$ and $K \le 19.75$)
 derived by Beckwith et al. (1998),
 the discovery of the possible dusty starburst galaxy merger  HR 10 implies
 that the fraction is  likely to be fairly large.
 Furthermore, the increase of number fraction of apparently interacting/merging galaxies
 with redshift in the HDF (Driver et al. 1998), the very small fraction (less than 8 \%)
 of early-type E/S0 galaxies in the HDF (Marleau \& Simard 1998),
 and the observed high surface density of EROs (Beckwith et al. 1998; Dey et al. 1998) 
 combine to motivate  us to interpret that a sizable fraction of the present-day
 elliptical galaxies were previously 
 ongoing major mergers with dusty starburst at intermediate and high  redshift.
 However,  spectroscopic studies of EROs have not been so accumulated yet which
 can reveal unambiguously the redshift of EROs thus determine whether EROs are really
 dusty starburst galaxies or such   high redshift red elliptical galaxies as have been 
 recently discovered by the VLT (Ben\'\i tez et al. 1998).
 Future extensive spectrophotometric studies which can reveal 
 whether the extremely red colors of EROs result from dust effects or from galaxy  aging
 will determine the relative importance of major galaxy merging
 in the formation of high redshift EROs.

\acknowledgments
 We are  grateful to the anonymous referee for valuable comments, 
which contribute to improve the present paper.
K.B. and Y.S. thank to the Japan Society for Promotion of Science (JSPS) 
Research Fellowships for Young Scientist.

\clearpage

\figcaption{
Upper panel shows the time evolution of star formation rate  
of a dusty gas-rich major galaxy merger at z=1.0. 
The star formation rate of the merger becomes maximum ($\sim 378 \rm  M_{\odot}/$yr)
at the time $T=1.3$ Gyr.
Lower panel shows the  rest-frame SED of the  galaxy merger at $T=1.25$ (dotted line),
$T=1.3$ (thick solid), and $T=1.35$ Gyr (dashed).
For comparison, the SED of the merger without dust extinction and re-emission
is also given by thin solid line.
Note that the far-infrared flux of the merger rapidly increases  between $T=1.25$ and $T=1.3$ Gyr
and decreases between $T=1.3$ and $T=1.35$ Gyr.
Furthermore we can clearly see the effects of dust extinction and re-emission on the SED shape
in the  merger at $T=1.3$ Gyr by comparing the thick solid line
and the thin solid  one. 
In this figure, the observed SED of Arp 220 by Rigopoulou, Lawrence, \& Rowan-Robinson (1996) 
is also given by a thin solid line with open squares. The reason for our failure to
reproduce the observed SED around $10^{5} \rm \AA$  is essentially  that we do not
include the effects of small grains in the present study.
\label{fig-1}}

\figcaption{
Mass distribution of a galaxy merger at $T=1.3$ Gyr corresponding
to the epoch of maximum secondary starburst of the merger for total components (upper left),
old stellar components initially located in two disks (upper right), 
gaseous ones (lower left), and new stellar ones formed by secondary starburst (lower right). 
Each of the four frames  measures 64.4  kpc on a side.
\label{fig-2}}

\figcaption{
The synthesized image  of  a galaxy merger
at the redshift z=0.4, 1.0, and 1.5 
projected onto the $x-y$ plane
corresponding to the orbital plane of the  merger
for the HST WFPC2 (upper four panels) and the HST NIC2 (lower
four panels).
Here the morphology of the merger  at the epoch of the maximum starburst ($T=1.3$ Gyr)
is described.
The redshift of the merger is indicated in the upper right-hand corner of each panel
and a bar with the length of 10 kpc is superimposed in the lower right-hand corner.
Except for the extreme right panel, the exposure time for each  synthesized morphology 
is set to be ${10}^{4}$ sec (2000sec $\times5$) both for
the HST WFPC2 and the HST NIC2.
The exposure time of the synthesized optical and near-infrared
morphology of the merger in the extreme right panel is $1.24  \times {10}^{5}$ sec 
(2000sec $\times 62$) corresponding
to that of the Hubble Deep Field (HDF). 
The filter adopted  in the present study is F814W for the WFPC2 and F160W for the NIC2.
Each frame measures ${9.}^{\prime\prime}$9 on a side and
one pixel size is ${0.}^{\prime\prime}$1 for the WFPC2 and ${0.}^{\prime\prime}$076  for the NIC2. 
The raw image is convolved with a WFPC2 point-spread function (PSF) generated 
by Tiny Tim ver-4.4 (Krist \& Hook 1997) and 
the STSDAS FCONVOLVE task to convolve the PSF is used.
A  magnitude zero point   is 21.60 mag for the WFPC2 (Holtzman et al. 1995)
and 25.68 mag for the NIC2 (MacKenty et al. 1997).
We assume that a  sky surface brightness at $V$ band is 22.9 $\rm mag / {\rm arcsec}^{2}$ 
corresponding to 0.054  $e^{-}/\rm pixel/\rm sec$ (Biretta 1996).
Readout noise, gain, dark noise are set to be
5.2 (35) $e^{-}/\rm pixel/\rm readout$,
7.0 (5.4)  $e^{-}/\rm  DN$, and 0.005 (0.05) $e^{-}/\rm sec$, respectively, for
the HST WFPC2 (the HST NIC2). We adopted these values, based on the WFPC2 Instrumental
hand book ver-4.0 (Biretta 1996) and HST NICMOS handbook (MacKenty et al. 1997).
We used the  NKNOISE package in IRAF
in order to add noise to the images.
We did not consider  other noise sources such as 
cosmic rays and image anomaly  
in the present study. 
\label{fig-3}}


\end{document}